# Pd/Cu single atom alloys for catalysis: from single crystalline to nanostructured model systems for highly selective butanol dehydrogenation

Philipp Haugg[1], Jan Smyczek[1], Carsten Schröder[1], Paul Fröhlich[1], Paul Kohlmorgen[1], Stephan Appelfeller[2], Konstantin Neyman[3]* and Swetlana Schauermann[1,4]*

[1]Institute of Physical Chemistry, Christian-Albrechts-University Kiel, Max-Eyth-Str. 1, 24118 Kiel, Germany.
[2]MAX IV Laboratory, Lund University, 22100 Lund, Sweden
[3] Departament de Ciència de Materials i Química Física & Institut de Química Teòrica i Computacional (IQTC-UB), Universitat de Barcelona, Barcelona 08028, Spain, ICREA (Institució Catalana de Recerca i Estudis Avançats), Barcelona 08010, Spain
[4]Kiel Nano, Surface and Interface Science, Christian-Albrechts-University Kiel, 24118 Kiel, Germany.

*schauermann@pctc.uni-kiel.de
*konstantin.neyman@icrea.cat

**ABSTRACT**

Heterogeneous catalysts based on Single Atom Alloys (SAA) play a significant role in numerous technical processes. The fundamental working principles of these systems, however, remain poorly understood, especially the aspects related to the nanoscopic nature of bimetallic particles and the associated structure-reactivity relationships. In this study, we developed for the first time well-defined SAA catalysts consisting of Pd atomically dispersed in Cu nanoparticles prepared under ultra-high vacuum (UHV) conditions on model $Al_2O_3$/NiAl(110) support. Employing a unique combination of surface sensitive techniques – scanning tunneling microscopy (STM), infrared reflection absorption spectroscopy (IRAS), molecular beams – and density functional theory (DFT) calculations, we performed detailed structural characterization of these systems at the microscopic level. We demonstrate that Pd disperses atomically in Cu nanoparticles and becomes partly negatively charged. Importantly, these Pd/Cu nanostructured systems show an outstanding catalytic performance in selective dehydrogenation of butanol and exhibit 100 % selectivity toward butanal over a broad range of Pd loadings – the property that cannot be reproduced employing simplified single crystalline Pd/Cu(111) counterparts. The developed approach for preparation and characterization of these nanostructured SAA-catalysts lays a foundation for further fundamental-level catalytic studies on this important class of materials and their rational design for practical applications.

**INTRODUCTION**

One of the major challenges in the modern heterogeneous catalysis is a precise control over the selectivity of a catalytic process, which allows to preferentially produce one of many other thermodynamically feasible products and by this drastically reduce the energy amount and costs required for separation. Another important costs-related aspect is that many of the currently employed catalysts contain nanoparticles (NPs) consisting of highly expensive ultra-rare transition metals, such as Pd or Pt, which are only partly exposed to the reactants, while the majority of the atoms in the bulk do not participate in the reactions. An overarching solution to both problems could be the use of hybrid nanoparticles based on Single Atom Alloy (SAA) systems, in which catalytically highly active ultra-rare elements (Pd, Pt) are atomically dispersed in more inert yet normally more selective host metals such as Cu, Ag or Au.[1-8] The major advantage of such hybrid systems is the possibility to design a catalytic surface, in which two different types of catalytically active sites can be independently tuned, and by this the selectivity and the overall activity can be precisely controlled. Also formation of extended ensembles of active atoms responsible for undesired full dissociation of hydrocarbons can be prevented.[9]

Several proof-of-principle studies show that this concept can be successfully applied in different types of heterogeneous catalysis including thermal, electro- and photocatalysis, as recently reviewed by Sykes[1,2], Friend[3], Zaera[5], Corma[6] and others.[4,7] Hydrogenation and oxidation of different hydrocarbon compounds over SAAs were reported both for powdered catalysts acting under realistic reaction conditions[4,6,7,10-12] and for model single crystalline SAAs investigated via rigorous surface science approach under controlled ultra-high vacuum (UHV) conditions.[1,3,5,13] In the latter studies, some industrially highly important reactions were studied on SAA systems consisting of an active compound embedded into a single crystalline host metal.[1,3,5,13] Despite the impressive developments in this field demonstrated both on powdered and on model single crystalline SAA-based catalysts, many aspects related to the structure-reactivity relationships and the reactions mechanisms remain poorly understood. This shortcoming is caused mainly by a vast complexity of the real powdered catalysts, which makes them inaccessible to accurate surface sensitive methods. On the other hand, the absolute majority of the reported mechanistic studies[4,6,7,10-12,14-18] was conducted on the simplified single crystalline SAA catalysts, which do not completely reproduce the nanoscopic nature of powdered catalysts, so that the direct transfer of the structure-reactivity relationships from the singe crystals to the nanostructured materials is not straightforward. Thus, the small size of SAA nanoparticles (NPs), their potentially altered electronic properties, the presence of low-coordinated sites, e.g. edges and corners and strong metal-support interaction might critically affect the activity and selectivity. To the best of our knowledge, the role of nanoscopic nature of SAA-based powdered catalysts was not addressed in the fundamental-level studies employing rigorous surface science approach so far.

To overcome this limitation, we developed for the first time a well-defined model system consisting of SAA-based nanoparticles supported on planar ultra-thin oxide film and studied their catalytic performance and the reaction mechanism for selective dehydrogenation of alcohols to ketones. This chemical process is gaining significant interest as a new emerging approach for reversible hydrogen storage.[3,4,6-8] There, the alcohol/ketone pairs can be utilized as the chemical storage media for controlled accumulation and release of molecular hydrogen by hydrogenation of ketones and dehydrogenation of alcohols. The major challenge associated with conversion of alcohols to ketones is their possible decomposition to carbon monoxide, alkenes and carbonaceous species, which can be potentially avoided on SAA-based catalysts.

For preparation of nanostructured SAA model catalyst, Pd was introduced as an active metal into the Cu nanoparticles that were prepared on top of planar well-defined $Al_2O_3$ thin film epitaxially grown on NiAl(110) single crystal. We performed for the first time a comprehensive structural characterization of these systems and addressed their reactivity by a unique combination of surface sensitive techniques including infrared reflection absorption spectroscopy (IRAS), scanning tunneling microscopy (STM), molecular beam techniques and temperature programed desorption (TPD) under well-defined UHV conditions. The specific focus was on a structural characterization of these materials via STM and IRAS by employing CO as a probe molecule for different adsorption sites. This combined approach allows a comprehensive description of these novel model systems, which cannot be obtained by a standard characterization via STM alone. Complementarily, density functional theory (DFT) calculations were performed on Pd embedded into Cu-NPs and Cu(111) single crystal to address structural and electronic properties of these materials. To investigate the effect of the nanoscopic nature of SAA-based Pd/Cu-NPs catalysts on the catalytic performance, decomposition of butanol to butanal was analyzed compared to Pd embedded into single crystalline Cu(111). Specifically, an outstanding catalytic performance was observed for nanostructured Pd/Cu-NPs catalysts, which exhibit high activity and 100 % selectivity toward butanal over a broad range of Pd loadings – the property that cannot be reproduced employing simplified single crystalline Pd/Cu(111) counterparts. Particularly missing undesired reaction – full decomposition to CO – suggests that the reaction-induced formation of Pd ensembles, which is required for multiple consecutives decomposition steps, can be efficiently prevented when Pd is embedded into Cu-NPs but not into Cu(111). Obtained results provide for the first time clear experimental evidence that nanostructured SAA-based catalysts exhibit specific properties due to their nanoscopic nature. The reported experimental approach for preparation and structural characterization of the nanostructured SAA-catalysts lays a foundation for fundamental catalytic studies on this important class of materials and holds a great potential for approaching their rational design.

**RESULTS AND DISCUSSION**

First, the preparation procedure of SAA-NPs supported on a model oxide film was developed. Figure 1 and 2 show the results of structural characterization of this model system by combination of IRAS and STM, respectively. CO was employed as a probe molecule, whose IR vibrational frequency sensitively depends on the chemical environment of Pd or Cu atoms. As an inert support, the single crystalline ultrathin $Al_2O_3$ oxide film was epitaxially grown on NiAl(110) according to a known procedure (Fig. 2b).[19] The STM (Fig. 2b) and LEED (Fig. S1 in Supporting Information, SI) images were found to be in agreement with the previously reported data. No metal sites of the underlying support are exposed in this film, so that it can serve as a non-reactive support in model catalytic studies.[19-21]

Next, Cu was deposited on $Al_2O_3$/NiAl(110) by physical vapor deposition (PVD) at elevated surface temperatures to ensure high mobility of Cu atoms and to achieve high degree of crystallinity of the resulting Cu-NPs. Two parameters were systematically tuned: *(i)* the amount of Cu, which was varied from 0.125 to 2 monolayers (ML), with ML being calibrated by quartz microbalance and STM; and *(ii)* the surface temperature during deposition (100 – 550 K) and the annealing temperature (up to 600 K), affecting the nucleation behavior and the crystallinity of the resulting NPs. Specifically, deposition of Cu at 400 K was found to produce stable Cu-NPs, which do not structurally change during further annealing to 500 – 600 K. In the last step, Pd was deposited onto Cu-NPs/$Al_2O_3$/NiAl(110) catalyst at different amounts (0.01 – 0.55 ML) and was annealed at systematically varying temperatures (300- 550 K) to allow for Pd/Cu single atom alloy formation. As a reference, single crystalline Pd/Cu(111) SAA-systems with varying Pd loading were prepared in an identical way.

Figure 1 shows a comparison of IR spectra obtained after CO adsorption on different single crystalline surfaces - Pd/Cu(111), Cu(111), Pd(111) (Fig. 1a) - vs. nanostructured materials Pd/Cu-NPs/$Al_2O_3$ (Fig. 1b). On Cu(111), CO adsorption at 100 K gives rise to a vibrational band at 2072 $cm^{-1}$ (spectrum 1a) previously attributed to CO adsorbed in on-top configuration on Cu(111).[22] Since this band lies in the same vibrational range as CO adsorbed in on-top configuration on Pd[23, 24], it appears useful to investigate SAA-catalysts not at 100 K, but at a higher temperature (≥ 230 K), at which CO does not adsorb on Cu[25-27] (see spectrum 1b, CO/Cu(111)) but is still present on Pd.[22-24] On Pd(111), CO is present at 230 K (spectrum 1c) and occupies the threefold-hollow (1953 $cm^{-1}$) and on-top (2088 $cm^{-1}$) Pd sites.[23, 24] Based on these results, we have chosen the temperature 230 K, at which adsorption of CO can be investigated on Pd embedded into Cu but not on Cu itself to avoid the overlap of Pd and Cu-related bands. Table S1 provides further information on CO vibrational frequencies previously reported on all relevant surfaces.

Next, 0.3 ML Pd was deposited onto Cu(111) at 300 K and the CO spectrum was obtained right after preparation (spectrum 1d) without annealing ("as deposited"). Pd was found to form small clusters, as evidenced by the appearance of the band at 1965 $cm^{-1}$ typical for threefold-hollow sites on small Pd-NPs.[19-21, 23, 24, 28, 29] The second band observed on this system at 2064 $cm^{-1}$ can be attributed to CO adsorbed on Pd in the on-top configuration.[19-21, 23, 24, 28, 29] Generally, the latter band might be related both to dispersed Pd atoms residing directly on Cu(111) as well as to on-top Pd sites on small Pd clusters. Interestingly, the band at 2064 cm$^{-1}$ appearing on "as deposited" Pd/Cu(111) has a substantially lower frequency than the on-top CO mode on Pd(111) (2088 cm$^{-1}$), most likely due to different chemical environment of Pd atoms, either in small Pd clusters and/or Pd embedded into Cu(111).

Next, this surface was annealed stepwise to different temperatures (500 – 800 K) and the CO spectra were obtained on these surfaces at 230 K (spectra 1e – 1h). After annealing to 500 – 600 K, the band at 1965 cm$^{-1}$ (CO at threefold-hollow Pd sites) vanishes, while the band related to the on-top adsorption configuration notably grows in intensity and shifts from 2065 to 2058 cm$^{-1}$. Both latter observations suggest that small Pd clusters disintegrate upon annealing and Pd becomes embedded into Cu(111) to form single atom alloy Pd/Cu(111), in agreement with the STM studies by Sykes et al.[1, 12] We attribute the newly formed band at 2058 cm$^{-1}$ to CO adsorbed on the single Pd atoms embedded into Cu(111), in line with the CO frequency (2064 cm$^{-1}$) reported by Trenary[30] for the SAA-Pd/Cu(111) system. The intensity of the band at 2058 cm$^{-1}$ starts to decrease at the annealing temperature of 700 K (spectrum 1g) and vanishes after annealing to 800 K (spectrum 1h), suggesting that Pd atoms diffuse into Cu bulk above 700 K.

Having determined the characteristic vibrational frequency of CO adsorbed on single Pd atoms embedded into Cu(111) as 2058 cm$^{-1}$, we investigated the adsorption sites available on nanostructured Pd/Cu-NPs surfaces (Fig. 1b). First, it was verified that CO does not adsorb on pristine Al$_2$O$_3$/NiAl(110) at the lowest possible temperature 100 K (spectrum 2), in line with previous reports.[19-21] Cu-NPs were grown on Al$_2$O$_3$/NiAl(110) at 400 K; the series of five spectra (spectra 2a) were recorded on these Cu-NPs at 100 K at increasing CO coverages. CO bands appearing here close to 2100 cm$^{-1}$ (2071 – 2106 cm$^{-1}$) suggest that CO adsorbs also in the on-top configuration.[22, 28] Based on the previous reports addressing CO adsorption at differently terminated Cu surfaces, we attribute the bands at 2088-2071 cm$^{-1}$ to CO adsorbed on Cu(111)[22, 31-34] the band at 2092 cm$^{-1}$ to Cu(110)[31, 35, 36] and the band at 2106 cm$^{-1}$ to the low-coordinated sites on Cu-NPs, such as e.g. edges and corners.[22, 28, 29, 32, 36] Note that the relative intensities of the bands might not directly reflect the absolute abundance of the surface sites due to dipole coupling and intensity borrowing effects.[37, 38]

Similarly to Cu(111), all CO bands vanish when the Cu-NPs are heated to 230 K (spectrum 2b). All other IR spectra obtained on the Pd-containing systems were therefore measured at 230 K to avoid the overlap of the Cu- and Pd-related bands and obtain the spectroscopic information only from Pd sites.

Next, we deposited small amount of Pd (0.05 ML) directly on Al$_2$O$_3$ in absence of Cu-NPs to address feasible formation of small Pd clusters on alumina support and their fate after annealing to 500 K. The spectra 2c and 2d were measured for CO adsorbed on this Pd/Al$_2$O$_3$ surface directly after Pd deposition and after annealing to 500 K, correspondingly. When adsorbed on freshly deposited Pd, CO shows two bands characteristic of threefold-hollow (1944 cm$^{-1}$) and on-top (2097 cm$^{-1}$) suggesting formation of small Pd clusters. Previously, similar behavior was observed upon CO adsorption on small Pd clusters.[19-21] Importantly, when this surface was heated to 500 K, both bands disappear (spectrum 2d), suggesting that at this temperature Pd atoms become sufficiently mobile and penetrate through the ultrathin alumina film into the bulk of the NiAl crystal. This phenomenon was previously described specifically for Pd diffusion through the ultrathin Al$_2$O$_3$/NiAl(110) film.[39]

Finally, we deposited the same amount of Pd (0.05 ML) on the supported Cu-NPs (1 ML), which were prepared in the same way as those shown in the spectra series 2a. The spectra 2e and 2f show CO adsorbed on this Pd/Cu-NPs system measured directly after Pd deposition (spectrum 2e, "as deposited") and after annealing to 500 K (spectrum 2f). On the "as deposited" system, the most prominent peak evolves at 2058 cm$^{-1}$, which is exactly the same frequency as detected on the Pd atoms atomically dispersed on Cu(111) (spectra 1f – 1g). There is also a small peak at 1944 cm$^{-1}$ related to CO adsorbed at the threefold-hollow sites of Pd clusters.[20, 21] Thus, directly after Pd deposition some Pd is agglomerated in Pd clusters residing on Al$_2$O$_3$ or Cu-NPs. However, after annealing to 500 K (spectrum 2f), the band at 1944 cm$^{-1}$ vanishes and only the intense band at 2058 cm$^{-1}$ remains preserved, which exhibits slightly higher intensity as compared to the "as deposited" Pd/Cu-NPs. With this, it can be safely concluded that after annealing to 500 K Pd clusters are present neither on Cu-NPs nor on the oxide support. Their fate might be twofold: *(i)* these Pd atoms become integrated into the Cu-NPs to form a SAA Pd/Cu-NPs as evidenced by the formation of a more intense band at 2058 cm$^{-1}$, and/or *(ii)* they diffuse through the ultrathin alumina film into NiAl bulk.

Thus, the developed preparation procedure results in SAA Pd/Cu-NPs supported on Al$_2$O$_3$/NiAl(110). These nanoparticles remain thermally stable up to at least 500 K, so they can serve as model nanostructured SAA catalysts. It should be emphasized that by employing CO as a highly sensitive probe molecule for Pd atoms in different chemical and structural environments, we can precisely follow incorporation of Pd into Cu(111) and Cu-NPs and formation of SAA nanoparticles.

The real space structural information on SAA-based Pd/Cu-NPs/Al$_2$O$_3$/NiAl(110) catalyst was obtained by STM. Figure 2a shows the distribution of single Pd atoms embedded into Cu(111) single crystal after annealing to 550 K, in agreement with previous studies[12, 14, 40] (see also the Figure S2 in SI and related discussion). While STM is capable of resolving individual Pd and Cu atoms in Pd/Cu(111) single crystal, it turned out to be impossible to achieve the atomic resolution on nanostructured systems. Figures 2c and d show the STM images obtained on the surface containing pristine Cu-NPs, which exhibit a clearly triangular form, suggesting the hexagonal symmetry of the topmost facet, i.e. (111) termination. The particles homogeneously cover the support with the density $7 \pm 0.6$ particles per 100·nm$^2$, the average particles size 3 nm or 460 Cu atoms per particle. Figure 2e, f shows this system after 0.05 ML Pd was deposited on the Cu-NPs at 300 K, while Figures 2g and h display the STM images obtained on Pd/Cu-NPs after annealing to 550 K. Both systems were prepared as described above and related to the spectra 2e and 2f shown in Figure 1b, correspondingly. It can be recognized that the particles do not noticeably change after Pd deposition and annealing: in both cases a fraction of the particles attains a somewhat rounder shape, while the rest remains triangular. Neither the particle density, nor average size noticeably changes. No clearly distinguishable single Pd atoms could be resolved on Cu-NPs so far, even though the spectroscopic data (Fig. 1b) show unambiguously that Pd is incorporated into the Cu-NPs. Thus, the capability of STM to resolve single metal atoms on relatively small nanoparticles is strongly limited and the comprehensive structural characterization of these systems requires a combination of microscopic (STM) and spectroscopic (IRAS) techniques that was for the first time realized in our study.

A very important observation can be made based on the IR data: the vibrational frequency of CO adsorbed at SAA-Pd/Cu alloy (2058 cm$^{-1}$) – both on Pd/Cu(111) and Pd/Cu-NPs – is significantly lower than the CO frequencies observed on both Pd(111) and Cu(111) surfaces (2088 and 2072 cm$^{-1}$, respectively). With this, the frequency of CO adsorbed on single Pd atoms embedded into Cu does not fall linearly between those detected for pure Cu and Pd. Generally, the origin of this effect can be twofold: first, the electronic structure of Pd embedded into Cu changes, which affects the distribution of the electron density on the CO molecular and results in the frequency shift. Second, the geometric effects, e.g. displacement of the foreign Pd atom from the regular position in the Cu lattice due to different spacial requirements, may cause a different bonding situation for CO and thus affect its vibrational frequency.

To clarify the origin of this effect, we performed DFT calculations using the plane-wave VASP code[41,42] for the single crystalline Pd/Cu(111) and nanostructured Pd/Cu-NPs model systems as well as for the reference monometallic systems Cu(111), Pd(111), Cu-NPs and Pd-NPs (see SI, Chapter 5 for details). Briefly, ca. 2 nm large 328-atomic Cu, Pd or Pd/Cu particles cut from the *fcc* metal structures and fully relaxed locally before and after CO adsorption were used as NP models. The single crystalline surfaces were modelled by six-layer slabs capable to accommodate several adsorbed CO molecules per supercell. Figure 3 summarizes key values calculated for Pd/Cu systems; see more detailed data and discussion in SI. Figure 3a shows exemplarily the calculated $Pd_2Cu_{326}$ NP with two single Pd atoms embedded into Cu-NP on the (111) terrace and at the corner. The position of the Pd atoms is indicated by the coordination number (CN), which counts the number of the nearest neighboring Cu atoms. Pd atom integrated in the Cu(111) facet is surrounded by 9 Cu atoms (CN=9) and Pd atom embedded into the corner has 6 Cu nearest neighbors (CN=6). The intermediate CNs 7 and 8 label Pd atoms in edges and (100) terrace, respectively. Figure 3b-d summarizes the following calculated parameters for two Pd coordination numbers (CN= 9 and 6): *(i)* the nearest distance between a metal atom (M=Pd, Cu) and CO, r(**OC-M**), *(ii)* the lateral distance between that metal atom and its nearest neighbors, r(Pd-Pd), r(Cu-Cu) and r(Pd-Cu), *(iii)* the partial Bader charge on a given metal atom, *(iv)* the calculated CO vibrational frequencies $\nu(CO)_{th}$, which are shown next to the experimentally measured frequencies $\nu(CO)_{exp}$, and *(v)* the CO adsorption energies $E_{ad}(CO)$. Note, that experimentally we cannot determine whether CO is adsorbed at a terrace or a corner, therefore the same experimental value $\nu(CO)_{exp}$ is provided for both cases CN= 9 and 6.

The experimentally observed trend – non-linear scaling of the CO vibrational frequencies for SAA-based systems – was generally supported by our calculations. Indeed, the frequency of CO adsorbed atop Pd atom in Cu(111) is smaller than frequencies computed for pure Cu(111) and Pd(111) (Fig. 3b). Similarly, for corner Pd atom in Cu-NP (CN=6), the computed frequency was found to be smaller than those calculated for the corner sites of both pure Pd-NPs and pure Cu-NPs (Fig. 3d). The only systems, for which the non-linear CO frequency dependence is not fully reproduced, was the Pd/Cu-NPs with Pd residing at the Cu(111) terrace (Fig. 3c): here, the CO frequency is calculated smaller than on the Pd(111) facet of a Pd-NP, in line with the experiment, however, the computed frequency of CO adsorbed on the Cu(111) terrace of a Cu-NPs was found to be even smaller. Overall, the theoretical results largely reproduced the striking observation on the non-linear scaling of the experimental CO vibrational frequencies for most of the Pd/Cu SAAs. Even more important, they fully reproduced the other main experimentally observed trend: the calculated vibrational frequency of CO adsorbed on a Pd atom was always lower, when this Pd atom was embedded into Cu than when it was surrounded by Pd atoms.

To understand the observed non-linearity, we analyzed calculated electronic and geometric structure of single Pd atoms embedded into Cu and CO adsorption energies. Importantly, Pd atom embedded into Cu host becomes notably negatively charged. Calculated Bader charge on Pd is -0.36 |e| for Pd/Cu(111) and for Cu-NPs it ranges from -0.28 |e| for a corner Pd atom to -0.34 |e| for a Pd atom embedded into a (111) facet (see SI for other Pd positions). Similar Bader charge of -0.39 |e| on Pd embedded into Cu(111) was reported by Sakaki et al.[43] With this, our results suggest that the single crystalline and nanostructured Pd/Cu systems exhibit similar Bader charges of single Pd atoms irrespectively of their specific CN. Generally, alone the notable negative charge on Pd can explain the measured and calculated red shift of the CO vibrational frequency. In particular, the Blyholder model predicts that increased negative charge on the metal should result in a stronger back-donation of the electron density form the metal to the antibonding $\pi^*$ CO orbital, which weakens the CO bond and results in a red shift of the vibrational frequency,[44] as previously demonstrated, for instance, for CO adsorbed on negatively charged Au atoms.[45, 46] Oppositely, with increasing positive charge on the metal, vibrational frequency of adsorbed CO typically shifts to higher values.[47-49]

The computed CO adsorption energies do not show a non-linear dependence on the composition of the alloy - the values calculated for the alloys were found to lie in between those calculated for pure Cu and Pd surfaces, either single crystalline (Fig. 3b) or nanostructured (Fig. 3c and d). Thus, the change in the CO adsorption energy does not seem to play a determining role in the experimentally observed non-linearity of CO vibrational frequency. Our computational results also indicate that Pd/Cu alloys undergo substantial structural changes as compared to the pure metals, both for single crystalline and nanostructured materials. Indeed, Pd atoms in the Cu host experience compressive strain of around – 5%, with Pd–Cu distances shortened to 2.58 – 2.61 Å as compared to Pd–Pd distance 2.72 – 2.77 Å in Pd bulk or Pd-NPs. On the other side, the computed Pd–Cu distance is longer in the Pd/Cu alloys than the Cu–Cu distance 2.54 – 2.59 Å in bulk Cu or Cu-NPs. Notably, the Pd–CO bond length computed for Pd/Cu is systematically elongated by ~0.04 Å relative to that for pure Pd, independent of the coordination of the embedded Pd atom. Furthermore, in all investigated Pd/Cu alloys the Pd atoms with adsorbed CO are partially lifted by 0.05 – 0.24 Å above the Cu surface plane due to larger atomic radius of Pd (see Fig. S 3, Table S 3 and further discussion in Chapter 5 of SI).

Overall, the theoretical data reveal that electronic structure of SAA-based Pd/Cu surfaces – both single crystalline and nanostructured – undergo substantial perturbation with Pd atoms becoming negatively charged. This partial negative charge is most likely responsible for the lowering the vibrational frequency of CO adsorbed at a single Pd atom embedded into a Cu lattice as compared to single Pd atoms in the monometallic form. On all types of computed alloys, substantial structural changes were detected: *(i)* Pd-Cu distances within the surface plane elongated compared to the regular Cu-Cu lattice distance and *(ii)* embedded Pd atoms lifted above the surface plane of Cu. Both structural changes evolve most likely to reduce the strain resulting from incorporation of larger Pd atoms into a smaller Cu lattice.

Having addressed the structural properties of the SAA-based nanostructured model catalysts, we performed catalytic test measurements on selective dehydrogenation of butanol to butanal over Pd/Cu-NPs with a broad range of alloys compositions combining TPD and IRAS (Fig. 4-5 and Fig. S 4 in SI). Figure 4a shows the reaction scheme of two competing reaction pathways: *(i)* the target process of butanol dehydrogenation to butanal proceeding through the surface reaction intermediate (RI) butoxy species and *(ii)* the undesired reaction of complete butanol decomposition to CO. Figure 4b and c displays the evolution of two gaseous products – butanal vs. CO – over pristine Pd(111) and Cu(111) reference surfaces. While Cu(111) shows no activity towards both reaction pathways, Pd(111) catalyzes only full decomposition of butanol to CO. Also on the "as deposited" single crystalline Pd/Cu(111) alloy containing small Pd clusters (Fig. 4d), CO evolves as the only reaction product. Interestingly, CO desorbs from the Pd/Cu(111) surface at lower temperature than from pristine Pd(111) (380 K on alloy vs. 485 K on pure Pd). In contrast, when the Pd/Cu(111) alloy was annealed to 550 K and single Pd sites were formed (Fig. 4e), the target product butanal was detected, while decomposition to CO was substantially suppressed, however, not completely prevented. This trend can be rationalized based on the population of Pd aggregates comprising multiple Pd atoms: while on the "as deposited" surface the Pd aggregates are populated and can efficiently catalyze multiple decomposition steps leading to CO formation, such multi-atomic aggregates nearly vanish after annealing, so that predominantly single Pd atoms embedded into Cu remain present on the surface. The latter species are likely not capable to strongly bind the reaction intermediate and conduct multiple C-H and C-C bond cleavage steps, which leads to formation of butanal.

Importantly, Pd deposited on Cu-NPs/Al$_2$O$_3$ shows high activity and selectivity toward butanal even when this system was not annealed to 550 K. Figure 4f displays the formation rates of butanal and CO on the "as deposited" Pd/Cu-NPs/Al$_2$O$_3$, comprising both dispersed Pd atoms and small Pd clusters. For this system, both products form at substantial amounts, while the "as deposited" single crystalline Pd/Cu(111) counterpart was found to be active only toward CO formation under the identical conditions. After annealing, leading to Pd dispersion in Cu-NPs, the catalyst shows an outstanding performance (Fig. 4g): substantially higher butanal formation rate than on the annealed Pd/Cu(111) (Fig. 4e) and completely suppressed CO formation, resulting in 100% selectivity toward butanal. The overall quantity of butanal formed over annealed Pd/Cu-NPs exceeds by a factor of two the butanal amount formed over annealed single crystalline Pd/Cu(111) containing the same amount of deposited Pd atoms. By performing IR measurements, we were also able to detect the reaction intermediate – butoxy species – on Pd/Cu-NPs catalysts (see Chapter 6, Fig. S 4 in SI).

Having detected a superior catalytic performance of Pd/Cu-NPs, we conducted the reactivity studies on butanol dehydrogenation in a broad range of Pd coverages (0.01 – 0.55 ML). Figure 5 shows the butanal and CO yields obtained on Pd/Cu(111) (Fig. 5 a,b) and Pd/Cu-NPs (Fig. 5 d,e) annealed catalysts plotted as a function of Pd coverage. The yields of both products were calculated as an integral intensity below the TPD curves. Figure 5c,f displays the resulting selectivity towards butanal for both types of catalysts. In the entire Pd coverage range, Pd/Cu-NPs catalysts show substantially higher formation rate of butanal than Pd/Cu(111) single crystals. Importantly, full decomposition to CO is completely suppressed over Pd/Cu-NPs for all Pd coverages (Fig. 5e), while growing Pd loading on single crystalline alloys results in notably increased CO formation (Fig. 5b). The overall selectivity to butanal was found to be unity in the entire investigated range of Pd coverages on Pd/Cu-NPs, while it rapidly declines to 0.2 on the Pd/Cu(111) single crystalline alloys. The latter effect observed on Pd/Cu (111) is most likely associated with agglomeration of Pd atoms to larger species capable of multiple dissociation steps and resulting in CO formation.

Thus, the SAA-based catalysis consisting of Pd atoms embedded in Cu-NPs show an outstanding catalytic performance both in terms of the absolute amount of formed butanal and the selectivity as well as maintain these catalytic properties over large range of Pd coverages. In contrast, the single crystalline Pd/Cu(111) catalysts demonstrate quickly declining selectivity with growing Pd content and lower absolute activity. These observations must be related to the nanoscopic nature of Cu-NPs, in which Pd can be embedded not only into the regular Cu(111) terraces but also at the edges and corners, whose catalytic activity can be substantially different to the single crystalline Pd/Cu(111) surface. Further, Pd can potentially reside close to the boundary between Cu-NPs and the support and thus be strongly affected by strong metal-support interactions.

**CONCLUSIONS**

Summarizing, we developed for the first time well-defined SAA catalysts comprising Pd atomically dispersed in Cu-NPs that were prepared under controlled UHV conditions on model Al$_2$O$_3$/NiAl(110) support. Employing a powerful combination of surface sensitive techniques (IRAS, STM, TPD) and DFT calculations, we performed detailed structural characterization of these systems at the microscopic level and compared them with the simplified single crystalline Pd/Cu(111) model surfaces. Pd was found to disperse atomically in Cu-NPs and become partly negatively charged. Importantly, the nanostructured Pd/Cu systems were found to exhibit an outstanding catalytic performance in selective dehydrogenation of butanol to butanal with 100% selectivity for a broad range of Pd loadings. This catalytic property could not be reproduced over simplified single crystalline Pd/Cu(111) counterparts and must be related to the nanoscopic nature of Pd/Cu particles, e.g. the presence of low-coordinated site and/or altered electronic properties. Particularly, the completely missing undesired reaction – full decomposition to CO – suggests that the formation of Pd ensembles, which is required to perform multiple consecutives decomposition steps, can be efficiently prevented when Pd is embedded into Cu-NPs as opposite to Cu(111) single crystals. The reported approach for the preparation and structural characterization of these nanostructured Pd/Cu-NPs catalysts holds a great potential for designing a broad range of model SAA-based nanostructured materials and lays a foundation for further fundamental-level catalytic studies on this important class of materials and their rational design for practical applications.

## ACKNOWLEDGMENTS

Financial support by the Deutsche Forschungsgemeinschaft (DFG, German Research Foundation; Project-ID 565361343, SCHA 1477/9-1) is gratefully acknowledged. We acknowledge the MAX IV Laboratory for beam time on the FlexPES beamline under proposal 20241754. Research conducted at MAX IV, a Swedish national user facility, is supported by Vetenskapsrådet (Swedish Research Council, VR) under contract 2018-07152, Vinnova (Swedish Governmental Agency for Innovation Systems) under contract 2018-04969 and Formas under contract 2019-02496. KN thanks the *Agencia Estatal de Investigación* of the Spanish *Ministerio de Ciencia, Innovación y Universidades* (MICIU/AEI/10.13039/501100011033) and ERDF/EU for research funding through grants PID2021-128217NB-I00 and CEX2021-001202-M.

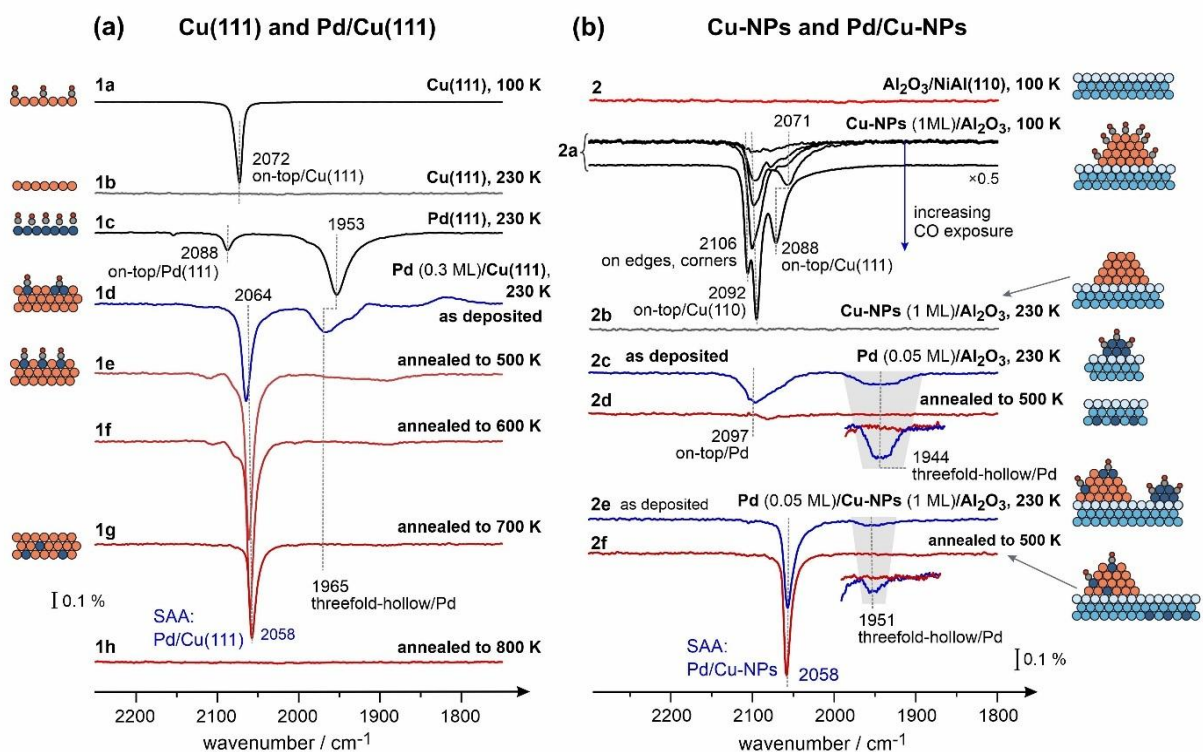

**Figure 1**: IR spectra of CO adsorbed as a probe molecule for different adsorption sites on (a) single crystalline and (b) nanostructured SAA Pd/Cu model catalysts and their precursors. The type of surface and the annealing temperatures are indicated next to the spectra. For detailed description see the text and SI, Chapter 2.

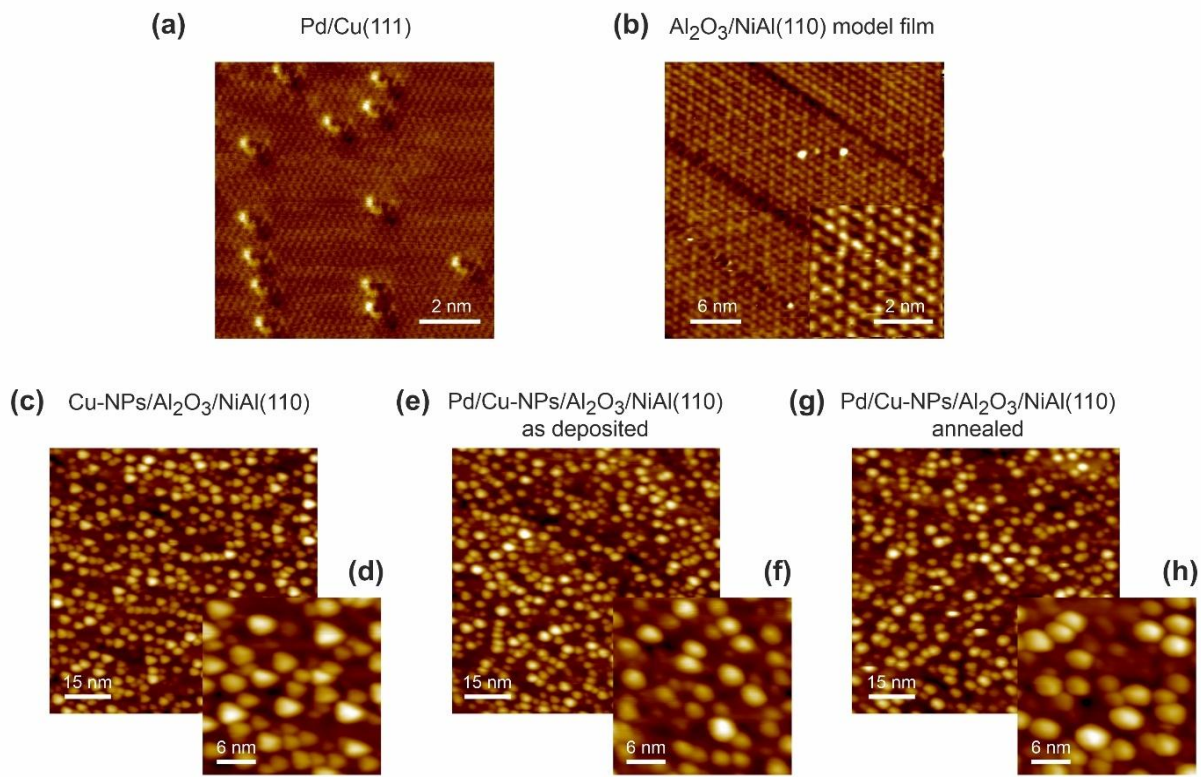

**Figure 2**: STM images of (a) SAA-Pd/Cu(111) alloy obtained after deposition of Pd on Cu(111) at 300 K and annealing to 550 K; (b) ultrathin $Al_2O_3$/NiAl(110) model support; (c-d) pristine Cu-NPs supported on $Al_2O_3$/NiAl(110); (e-f) "as deposited" Pd/Cu-NPs/$Al_2O_3$/NiAl(110) and (g-h) Pd/Cu-NPs/$Al_2O_3$/NiAl(110) annealed to 550 K (full details are in SI, Chapter 2).

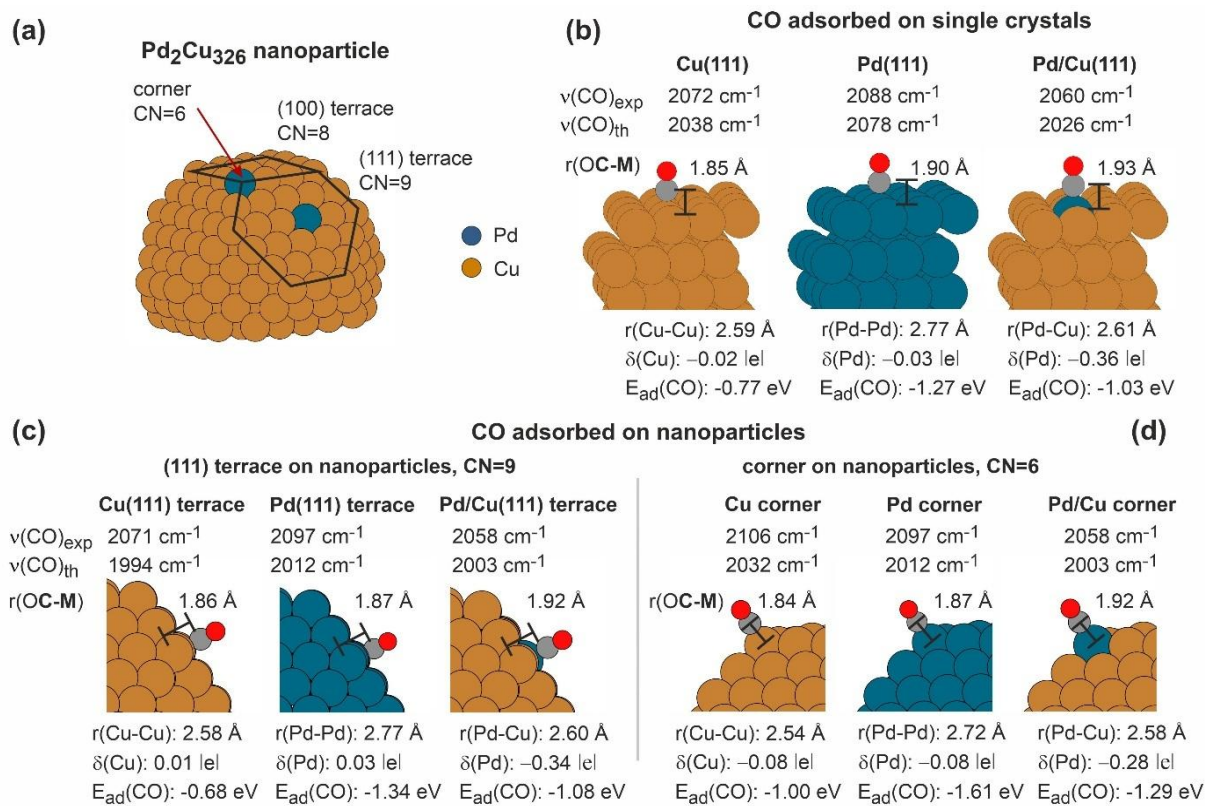

**Figure 3**. Results of DFT calculations performed on Pd-, Cu- and Pd/Cu-NPs (a, c, d) and their (111) single crystalline counterparts (b). The calculated M-CO and M-M interatomic distances, CO adsorption energies and vibrational frequencies as well as Bader charges of the atoms M are shown. See computational details and full discussion in SI, Chapters 1 and 5.

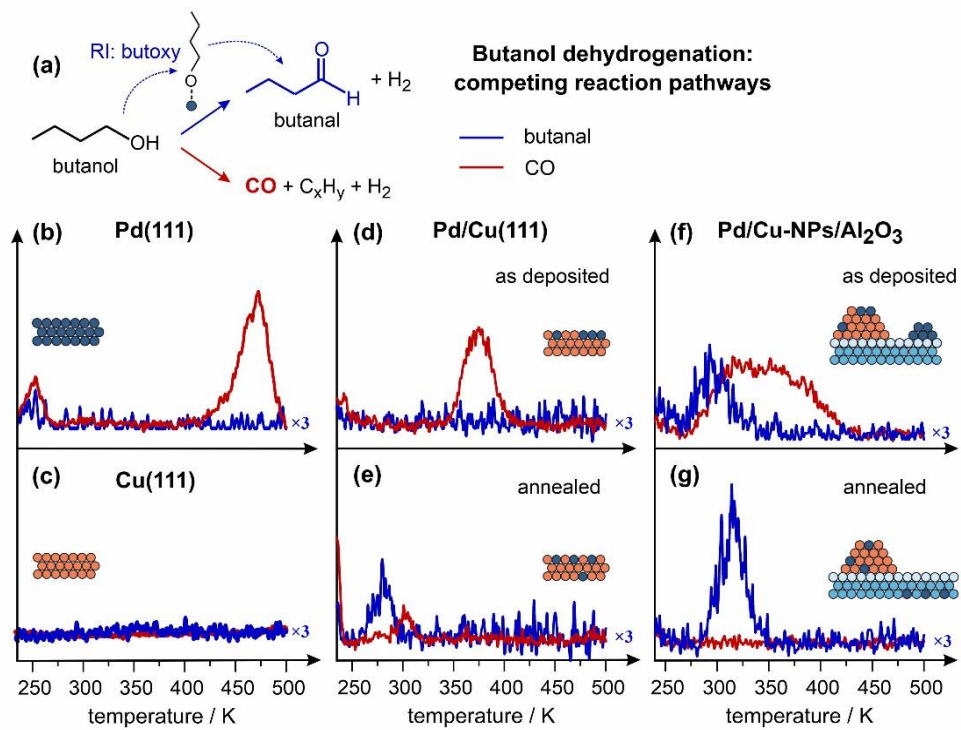

**Figure 4.** Reactivity of model catalysts in butanol dehydrogenation proceeding according to competing reaction pathways shown in (a). Formation rates (arbitrary units) of butanal (blue, multiplied by 3) and CO (red) obtained over Pd(111) (b), Cu(111) (c), and SAA-based catalysts: Pd (0.3 ML)/Cu(111) – "as deposited" (d) and annealed to 550 K (d); as well as Pd (0.3 ML)/Cu-NPs (1 ML)/Al$_2$O$_3$ – "as deposited" (f) and annealed to 550 K (g). Full details are in SI, Chapter 3.

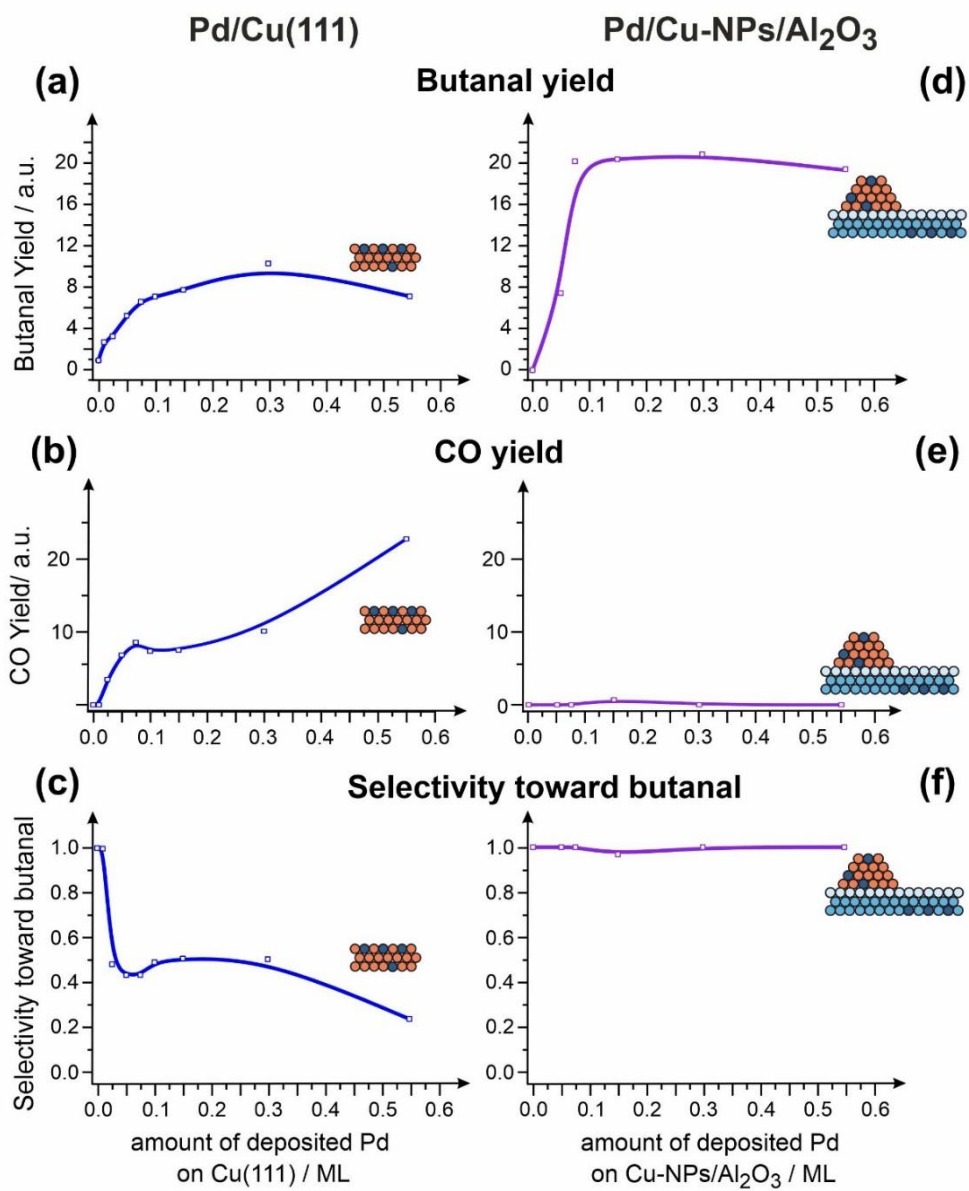

**Figure 5**. Catalytic performance of SAA Pd/Cu(111) (a,b,c) and Pd/Cu-NPs/Al$_2$O$_3$ (d,e,f) model catalysts (both annealed to 550 K) measured as a function of amount of deposited Pd. Shown are the butanal (a,d) and CO (b,e) yields as well as the resulting selectivity (c,f).

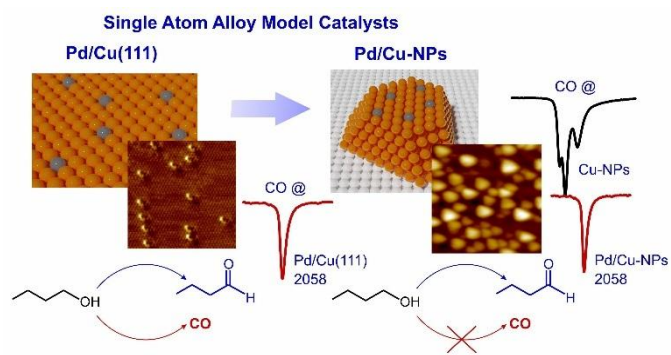

Table of Content (TOC)